\documentclass[aps,prl,twocolumn,showpacs,amsmath,amssymb]{revtex4}
\usepackage{graphicx}
\usepackage{dcolumn}
\usepackage{bm}
\begin{document}

\title{Controllable soliton emission from a Bose-Einstein condensate.}
\author{Mar\'{\i}a I. Rodas-Verde and Humberto Michinel}
\affiliation{\'Area de \'Optica, Facultade de Ciencias de Ourense,\\ 
Universidade de Vigo, As Lagoas s/n, Ourense, ES-32004 Spain.}

\author{V\'{\i}ctor M. P\'erez-Garc\'{\i}a}
\affiliation{Departamento de Matem\'aticas, E. T. S. I. Industriales,\\ 
Universidad de Castilla-La Mancha, 13071 Ciudad Real, Spain.}

\begin{abstract}
We demonstrate, through numerical simulations, the controllable emission of 
matter-wave bursts from a Bose-Einstein Condensate in a shallow optical dipole trap. 
The process is triggered  by spatial variations of the scattering length along the 
trapping axis. In our approach, the outcoupling mechanism 
are atom-atom interactions and thus, the trap remains unaltered. Once emitted, the matter wave 
forms a robust soliton. We calculate analytically the parameters for the experimental 
implementation of this atomic soliton {\em machine gun}.
\end{abstract}

\pacs{42.65.Jx, 42.65.Tg}

\maketitle
{\em Introduction.-}  After the remarkable experimental realization of Bose-Einstein 
condensates (BEC) with alkali atoms\cite{Anderson95} it was soon realized that the coherent 
behavior of the atom cloud could be used as the basis for an ``atom laser".
The first realization of such device used short radio-frequency pulses 
as an outcoupling mechanism, flipping the spins of some of the atoms and  
releasing them from  the trap \cite{Mewes97}. Later, other atom 
lasers were built with different configurations leading to pulsed, 
semi-continuous or single-atom coherent sources\cite{Bloch,Hagley,Bloch2,Pepe,Ultimo}.

There are important differences between these devices and photon pulsed lasers.
The nature of the waves and the lack of a population inversion mechanism 
are the most evident. However, from the practical point of view, one of the 
most remarkable is that a pulsed BEC laser will show a significant spreading 
of the atom cloud if the number of atoms per pulse is large. 

A way to overcome this difficulty would be  to use atomic bright solitons \cite{solitons1,solitons2} 
as matter wave pulses. This idea was explored in Ref. \cite{Carr04} to generate a train of 
solitons by the mechanism of modulational instability (MI) and was referred to as an atomic 
soliton laser. However, although one could extract a few coherent solitons from such a system 
this atomic soliton laser would be very limited since: (i) the final number of atoms per 
pulse after MI is only a small fraction of the initial number of atoms 
in the condensate due to collapse processes, (ii) the number of solitons generated is not 
large and half of them would be directed backward  (iii) the trap should be destroyed for 
outcoupling and (iv) the pulses will travel at different speeds once the trap is removed. 

Thus, it is important to discuss new outcoupling mechanisms for atom lasers. 
This is specially interesting since the techniques for generating BEC with growing number of particles 
and their physical properties are nowadays well established and the current 
challenges in the field face the applications of coherent matter waves to 
the design of practical devices \cite{chip}. 

In this letter, we show how a highly controllable train of up to several hundreds 
of matter-wave solitons can be extracted from a BEC {\em without altering the trap properties} 
but instead acting on the scattering length, by changing it along the atom cloud.

{\em System configuration and theoretical model.-} Let us assume that a large BEC is strongly trapped 
in the transverse directions ($x,y$) and weakly confined in the longitudinal 
one ($z$). We now consider the effect of a sharp variation along $z$ of the 
scattering length, which is changed from positive (or zero) to negative, thus 
making it inhomogeneous. This can be done with magnetic \cite{FB1} or optical\cite{FB2} 
techniques and therefore this region of negative scattering length can be displaced 
along the condensate. Thus, when it is located close to one edge of the trap, 
overlapping the wing of the cloud, the tail of the condensate may be able to 
form a single soliton which, because of its higher internal energy will be 
outcoupled from the cloud and thus emitted outward. When the condensate refills 
the gap left out by the outgoing pulse a new soliton would be emitted. This 
process would continue while there is a large enough remnant of atoms in the trap 
and would lead to a soliton burst escaping from the BEC.

Thus, we will describe our system  of $N$ weakly interacting bosons of mass 
$m$, trapped in a potential $V(\vec{r})$ is the mean field limit by a 
Gross-Pitaevskii equation of the form
\begin{equation}
\label{GPE}
i \hbar \frac{\partial \Psi}{\partial t} =
- \frac{\hbar^2}{2 m} \nabla^{2}\Psi +
V(\vec{r})\Psi + U(z)|\Psi|^2 \Psi,
\end{equation}
where $\Psi$ is the condensate wavefunction, $N = \int |\Psi|^2 \ d^3 \mathbf{r}$ The 
coefficient $U(z) = 4 \pi \hbar^2 a(z)/m$ characterizes the 2-body interaction and since 
we will consider spatially inhomogeneous systems it will be a function of $z$.

In this paper, we consider a BEC tightly confined in ($x,y$) by a harmonic potential 
$V_\perp$ and more relaxed along $z$ where there is and optical dipole trap 
$V_z$\cite{Stamper98,Martikainen99}. Thus, we have

\begin{equation}
V(\vec{r})=V_\perp+V_z=\frac{m\nu^2_\perp}{2} 
\left( x^2+y^2 \right)+V_0
\left[1-\exp\left(-\frac{z^2}{L^2} \right)\right],
\end{equation}
where $V_0$ is the depth of the shallow optical dipole potential and $L$ is the 
characteristic width of the trap. We will consider situations in which the ground 
state of the optical dipole trap is much larger than the ground state of the 
transverse harmonic potential.

\begin{figure}[htb]
{\centering \resizebox*{1\columnwidth}{!}{\includegraphics{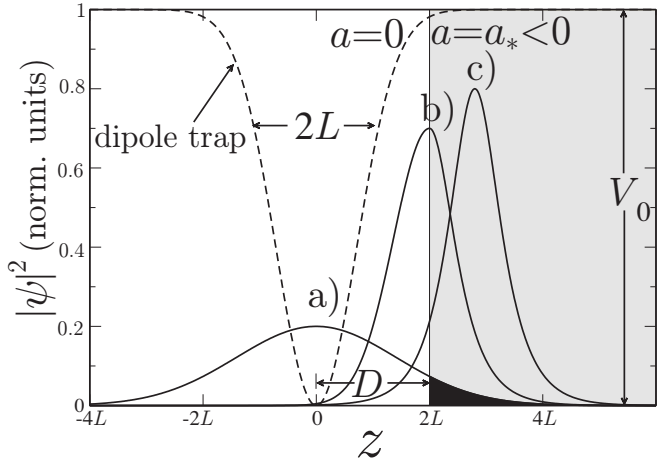}} \par}
\caption{Sketch of the system showing the axial trap (dashed line) and the shaded
region $z>D$ (=2$L$ in this particular case) which has negative scattering length $a$. 
For the rest of the cloud ($z<D$) $a$ is set to zero. The black zone displays the 
overlapping of the cloud and the region with negative $a$. This causes the outcoupling 
for a critical value of the scattering length $a_{cr}$. The continuous lines are the 
numerically calculated  profiles of eigenstates 
for a)  $a_*=0$ , b) $a_*=a_{cr}$ and c) $a_*=1.2a_{cr}$.}
\label{fig1}
\end{figure}

In Fig.\ref{fig1} we have plotted the geometry of the system 
The choice of a shallow Gaussian trap is very important for our model, since we 
are interested on studying the outcoupling of solitons along the $z$ axis. Thus, we 
need a potential barrier that can be overcome by the self-interaction effects. 
In this situation we can describe the dynamics of the condensate in the quasi-one dimensional
limit as given by a factorized wavefunction of the form \cite{Perezgarcia98} 
$\Psi(\boldsymbol{r},t)=\Phi_0(x,y)\cdot\psi(z,t)$, satisfying

\begin{equation}
\label{NLSE}
i\frac{\partial \psi}{\partial \tau} = - \frac{r_{\perp}^2}{2}
\frac{\partial^2\psi}{\partial z^2} + f(z)\psi + g |\psi|^2 \psi,
\end{equation}
where $r_\perp = \sqrt{\hbar/m\nu_\perp}$ is the transverse size of the cloud,
 $f(\eta) = V_z/(\hbar\nu_\perp/2)$, $\tau=\nu_\perp t$ is the time measured in units of the inverse of the 
radial trapping frequency
and $ g(z) =\sqrt{8}\pi r_\perp^2 a(z)$ is the effective interaction coefficient.

In Fig. \ref{fig1} we show a sketch of the setup to be considered in this paper. 
We show the axial optical dipole potential of depth $V_0$ and half width $L$. The shaded region ($z>D$)
has negative scattering length. A spatial variation along $z$ of the scattering length $a$, can 
be achieved by magnetically tunning the Feschbach resonances \cite{FB1} or by their optical 
manipulation by means of an additional laser beam \cite{FB2}. In this paper we choose a step
model for $a(z)$ in order to present the basic mechanism. We must stress that our ideas apply 
to smoother variations of $a$. Thus, we will use a dependence of the form
\begin{equation}
a(z) = \begin{cases} 0, & z< D \\
a_*<0, & z>D.
\end{cases}
\end{equation}
 The continuous lines in Fig. \ref{fig1} are the numerically calculated 
profiles of eigenstates of Eq. \eqref{NLSE} for three different values of $a_*$. 
The lower curve corresponds to he linear case ($a_*=0$), whereas 
lines b) and c) display the shape of the cloud for $a_*=a_{cr}$ and $a_*=1.2a_{cr}$, respectively.
The value $a_{cr}$ is the (negative) scattering length needed for emitting one soliton from the cloud. We will 
calculate it by means of an approximate analytical technique.

{\em Single soliton emission.-} In Fig. 1 (black region), it is shown the overlapping of
the tail of the trapped cloud and the zone where $a = a_*$. Due to the atom-atom interaction, 
the trapping changes in this region. Thus, if $a_*$ is negative enough, the minimum of the
effective potential will be displaced toward this region and the cloud will move to the 
shaded region of Fig. \ref{fig1}. 

\begin{figure}[htb]
{\centering \resizebox*{1\columnwidth}{!}{\includegraphics{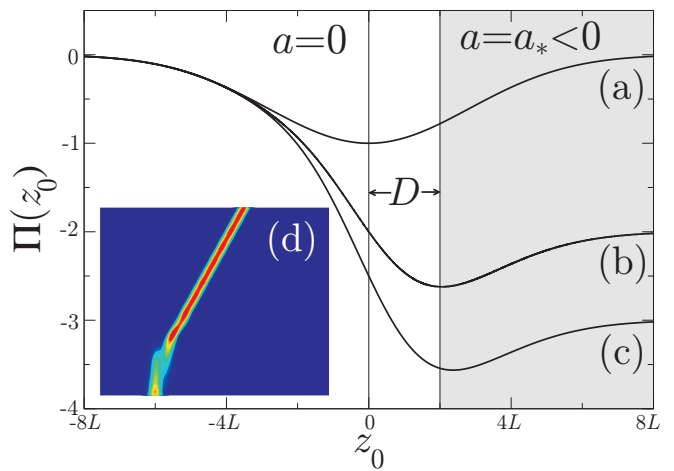}} \par}
\caption{[Color Online] Effective potentials corresponding to different values
of the scattering length: (a) $a_*=0$, (b) $a_*=a_{cr}$, (c) $a_*=1.2a_{cr}$. In (d) we show a 
pseudocolor plot of $|\psi(z,t)|^2$ showing the 
soliton emission for $a_*=1.2a_{cr}$. The parameter values are
$V_0=\hbar\nu_\perp/2$, $\nu_\perp=1KHz$, $L=4r_\perp$, $D=2L$, 
$w=5.4r_\perp$ and $N\cdot a_{cr}=-100\mu m$. The spatial window size is $100r_\perp$, 
and time goes from $0$ to $4000\nu_\perp^{-1}$.}
\label{fig2}
\end{figure}                                                                                                              

To study this possibility we will use an averaged Lagrangian formalism \cite{AL}
for studying localized solutions of Eq. (\ref{NLSE}). We will use a Gaussian ansatz 
\begin{equation}
\psi(z,\tau)=Ae^{-(z-z_0(\tau))^2/2w^2}e^{i\left(v(\tau)z\right)},
\end{equation}
where $z_0(\tau)$ accounts for the motion of the center the cloud with velocity 
proportional to $v$. The standard calculations lead to the equations
 \begin{subequations}
\begin{eqnarray}
\label{variacional}
\ddot{z_0} & = & -\frac{d\Pi}{dz_0},\\
\label{potential}
\Pi(z_0) & = & -\sqrt{\pi}r_\perp^2\left\{\frac{V_0}{\hbar\nu_\perp}
\left(1- \frac{e^{-\frac{z_0^2}{w^2+L^2}}}{\sqrt{1+\frac{w^2}{L^2}}}\right) \right .\nonumber  \\ 
  &  &  \left .+\frac{1}{\sqrt{8\pi}} \frac{aN}{w} \text{erfc}\left[\frac{\sqrt{2}\left(D-z_0\right)}{w}\right]\right\}, 
\end{eqnarray}
\end{subequations}
where $\text{erfc}(u)=\frac{2}{\sqrt{\pi}}\int_u^\infty \exp(-v^2)dv$ is the complementary 
error function. Thus, $z_0$ evolves like a classical particle under the 
effect of a potential $\Pi(z_0)$.This provides a qualitative understanding of the soliton emission:
for the linear case ($a=0$) the center of the cloud is located at the bottom of the Gaussian trap, 
which corresponds to its fundamental eigenstate [Fig. \ref{fig1} (a)]; as $a_*$ takes more 
negative values the effective trapping of the cloud is deformed and the minimum of 
the equivalent potential moves to the region with $z>0$. Fig. \ref{fig2} shows the equivalent 
potentials $\Pi$ given by Eq. (\ref{potential}) for different values of $a_*<0$.  
Fig. \ref{fig2}(a) shows the effective potential for $a_*=0$. As $a_*$ is decreased 
there is a limiting value $a_{cr}$ for which the potential $\Pi(0)=\Pi(\infty)$ [Fig. \ref{fig2}(b)], 
thus if the atom cloud is initially placed at $z_0 = 0$ it will oscillate around the minimum 
and escape $z_0(\tau) \rightarrow \infty$ for $\tau \rightarrow \infty$, 
a phenomenon which is called soliton emission\cite{emision}. 
The critical value of $a_*$ that corresponds to the threshold for soliton emission  
can be obtained within our formalism from the condition, $\Pi(0) = \Pi(\infty)$
which yields to:
\begin{subequations}
\begin{eqnarray}\label{acr}
Na_{cr} & = & \frac{\sqrt{8\pi}V_0}{\hbar\nu_\perp}\frac{wL}{\sqrt{L^2+w^2}}
\left[\text{erfc}\left(\frac{\sqrt{2}D}{w}\right)-2\right]^{-1} \\ & \approx & 
- \frac{\sqrt{8\pi}V_0}{\hbar\nu_\perp }\frac{wL}{\sqrt{L^2+w^2}}\approx5.0\Delta\frac{ L}{w},
\end{eqnarray}
\end{subequations}
the first approximation is valid for  $D \gtrsim w$ and the second for $V_0=\Delta\hbar\nu_\perp$, with $\Delta<1$, which 
provides a shallow trap with $w\gg L$.
We have found that the ratio between the exact numerical value and the prediction from 
Eq. (\ref{acr}) is a factor $1.15$ which is essentially due 
to the specific choice of the ansatz. In fact, once the soliton is emitted the 
wavefunction takes a hyperbolic secant shape and its width differs from the Gaussian by
a similar factor. However, taking a sech profile as variational ansatz does not yield to analytical 
results with the potentials we have considered.

For more negative values of $a < a_{cr}$ not only the soliton is outcoupled from the system but also it propagates
with a finite asymptotic speed in the nonlinear medium. An example of the effective potential for $a=1.2a_{cr}$ is shown 
in Fig. \ref{fig2}(c)  and the corresponding soliton 
emission in Fig. \ref{fig2}(d), for a $^7$Li condensate such as that of Ref. \cite{solitons1}

\begin{figure}[htb]
{\centering \resizebox*{1\columnwidth}{!}{\includegraphics{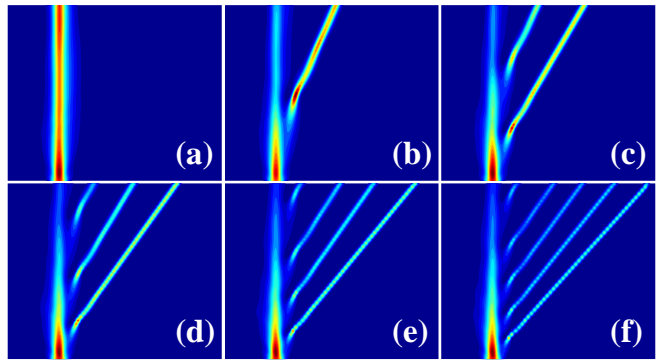}} \par}
\caption{\label{fig3}
[Color Online] Multiple emission of atomic solitons from a BEC
reservoir for different values of $a_*$: (a) $a_*=0.9a_{cr}$, (b) $a_*=2.0a_{cr} $, (c) $a_*=3.3a_{cr} $,
(d) $a_*=4.8a_{cr} $, (e) $a_*= 6.5a_{cr}$, (f) $a_*=8.7a_{cr}$.The vertical axis in each 
figure is time from $t=0$ to $t=1s$ and the horizontal width of each window is 
$100r_\perp\approx300\mu m$. The rest of the parameters are the same as for the 
inset of Fig. \ref{fig2}, except $D=2.5L$.}
\end{figure}

{\em Partial outcoupling and multisoliton emission.-} A deeper numerical exploration based on Eq. (\ref{NLSE}) reveals
more interesting effects which are beyond the averaged lagrangian description. An example is shown in Fig. 3 for parameter values
$V_0=\hbar\nu_\perp/2$, $\nu_\perp=1KHz$, 
$L=4r_\perp$, $D=2.5L$, $N=3\cdot10^5$, $w=5.4r_\perp$. 
The vertical axis in each figure is time from $t=0$ to $t=1s$ and the horizontal width of
each window is $100r_\perp\approx300\mu m$.  When  $a_*=0.9a_{cr}$ 
 the atom cloud  is only slightly deviated to the region with $a<0$ [Fig. \ref{fig3} (a) ]. Decreasing $a_*$
 down to $a_*=1.95a_{cr} $ leads 
  to the phenomenon of soliton emission with some remnant staying on 
the optical dipole trap [Fig. \ref{fig3}(b)]. However, as $a_*$ is made more and more negative we obtain the emission  
 of an integer number of solitons.

\begin{figure}[htb]
{\centering \resizebox*{1\columnwidth}{!}{\includegraphics{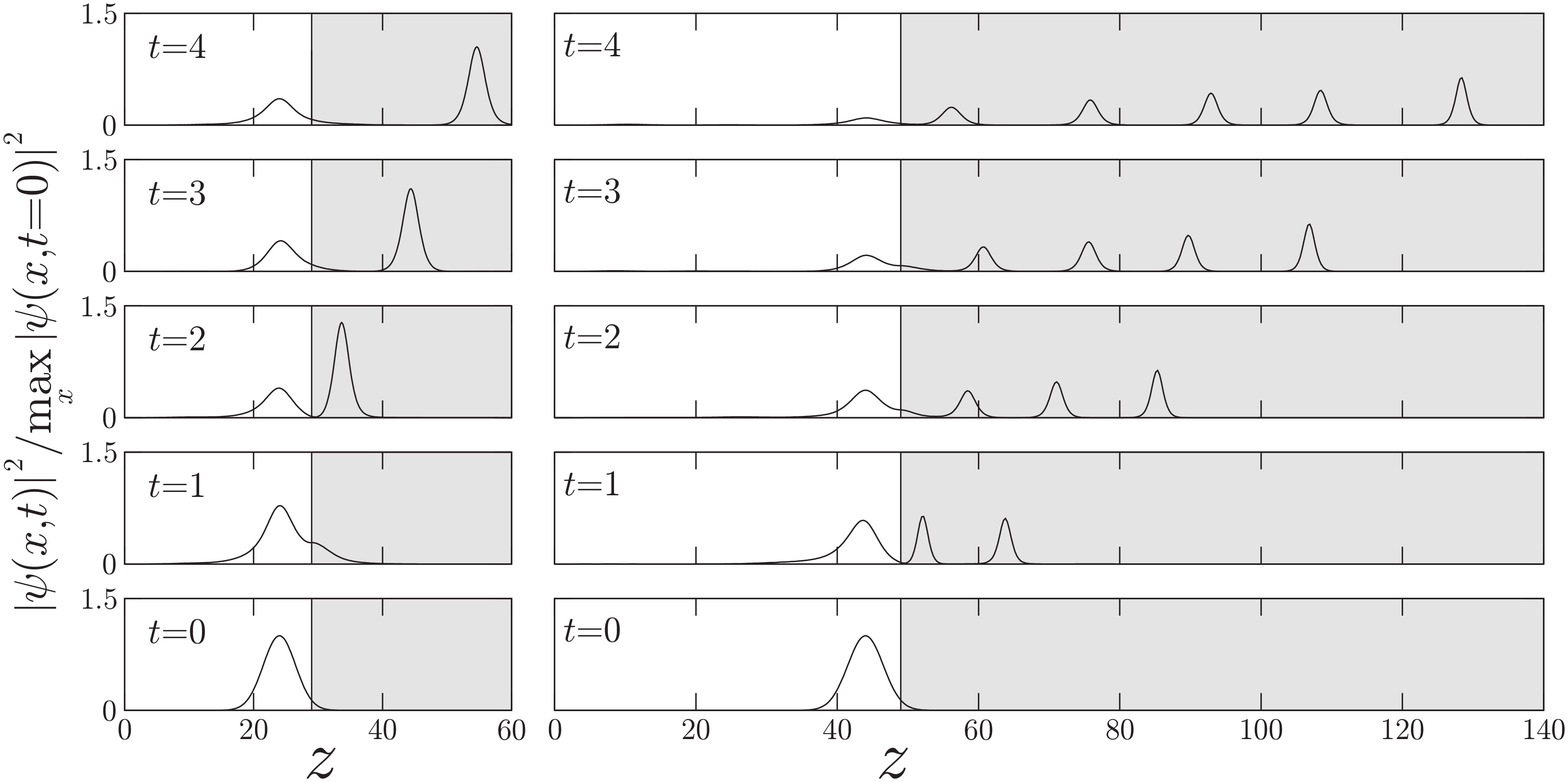}} \par}
\caption{\label{fig4}
Soliton emission for $D=2.5L$ and five different times from $0$ to $1s$.
Left: Emission of a single soliton for $a=1.1a_{cr}$. Right (b) Emission of 
five solitons for $a=4.4a_{cr}$.}
\end{figure}
                                                                                 
The phenomenon of partial emission can be qualitatively understood 
with the variational method. As it can be seen in Fig. \ref{fig2} the effective potentials
$\Pi$ have only one minimum for $D=2.0L$. However if the value of $D$ is increased to say 
$D=2.5L$, the potential $\Pi$ for values close to $a_{cr}$ presents one maximum between two
adjacent minima. This configuration of the effective potential causes the split of the 
matter wave in two parts: one remains in the trap and the other is emitted as a soliton.

\begin{figure}[htb]
{\centering \resizebox*{1\columnwidth}{!}{\includegraphics{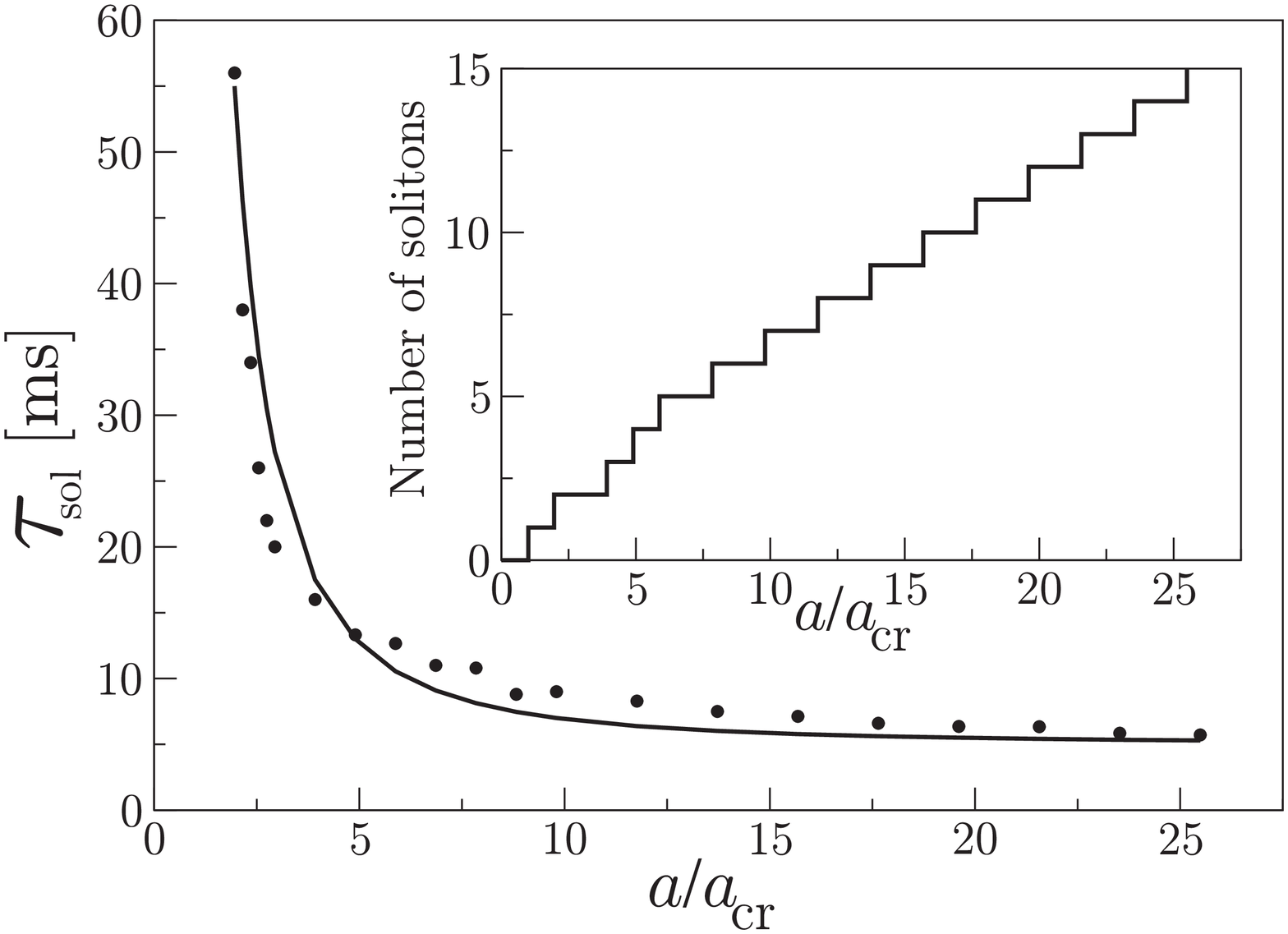}} \par}
\caption{\label{fig5}
Average time interval $\tau_{sol}$ between the emission of two solitons as a function of $a_*/a_{cr}$.
The curve follows a law of the type $\tau\propto a_*^{-2}$ (solid curve). Insert:
Number of solitons emitted as a function of  $a_*/a_{cr}$.}
\end{figure}

From our numerical exploration we have found some regularities 
that are noticeable. In first place, as it can be appreciated 
from Fig. \ref{fig3}, the velocities of the outgoing pulses are 
almost equal. Secondly, the intervals of emission are very regular and decrease with
the effective nonlinearity with a law of the form $\tau \sim (a_*/a_{cr})^{-2}$.
Finally, the values of the scattering length required for outcoupling several 
solitons are almost exact multiples of $a_{cr}N$ as it is clear by the 
approximate linear dependence of the number of solitons emitted in the
inset of Fig. \ref{fig5}. This can be interpreted as the requirement of a 
critical effective nonlinearity remaining in the trapped cloud for the formation of every soliton.
Finally, we must comment from Fig. \ref{fig4} that the amplitude of the emitted solitons
decays from the first one of the burst to the last approximately in a 30\%, depending on the 
number of pulses.

{\em Discussion and conclusions.-} These regular and controllable soliton trains could be 
realized with any atomic specie for which the scattering length could be made negative on a region of the space. 
This has been done with several atomic species such as 
$^7$Li, $^{85}$Rb and $^{133}$Cs. Taking for instance the experimental parameters of Ref. \cite{solitons1} for
 Lithium and assuming  $3\times 10^5$ atoms in the optical dipole trap we obtain that a few (2-3) solitons could be 
 emitted for a depth of $V_0 = \hbar \nu_{\perp}/2$. However,
 increasing the initial particle number in the zero-scattering length 
region could allow the observation of soliton trains with a higher number of solitons, e.g. for $3 \times 10^6$ we estimate up to $N_{sol} \sim 20$.
For $^{85}$Rb and taking parameter data from Ref. \cite{85Rb} we estimate about $25$. Even more interesting effects could be obtained 
with Cesium, for which experimental data of  Ref. \cite{133Cs} lead to an estimate of $N_{sol} \sim 200$. In all cases these numbers can be raised or lowered by acting 
on the parameters: $a_*, N, D$ and $V_0$.

In summary, we have proposed a novel mechanism for
outcoupling coherent matter wave pulses from a Bose-Einstein Condensate. Our system is able to 
perform a regular and controllable emission of atomic soliton bursts that are easily extracted
by an adequate choice of the control parameters. Using this mechanism a train of even several 
hundred of solitons could be coherently outcoupled from a condensate. As the techniques for coherently feeding 
the remaining condensate progress our idea could provide an outcoupling mechanism for a continuous 
atomic soliton laser.

This work was supported by Ministerio de Educaci\'on y Ciencia, Spain
(projects FIS2004-02466, BFM2003-02832, network
FIS2004-20188-E) and by Xunta de Galicia (project PGIDIT04TIC383001PR).


\end{document}